\documentclass[preprint,10pt]{elsarticle}

\usepackage{lineno, hyperref, subfig,
version, array, color}

\usepackage{amsmath}
\usepackage{amsbsy}
\usepackage{amssymb}
\usepackage{caption}
\usepackage{textcomp}
\usepackage{gensymb}
\usepackage{siunitx}
\usepackage{multirow}
\usepackage[table,xcdraw]{xcolor}
\usepackage{numcompress}
\usepackage{natbib}
\usepackage{har2nat}
\usepackage{color}
\usepackage{soul}
\usepackage[english]{babel}
\usepackage[utf8]{inputenc}
\usepackage{booktabs} 
\usepackage{longtable}
\usepackage[nomarkers, nolists]{endfloat}
\DeclareDelayedFloatFlavour*{longtable}{table}
\journal{}
\date{}

\begin{document}
\begin{frontmatter}

\title{\textit{In vitro} and sensory tests to design easy-to-swallow multi-particulate formulations}


\author[myfirstaddress]{Marco Marconati}

\author[mythirddaddress]{Felipe Lopez}
\author[mythirddaddress]{Catherine Tuleu}
\author[mythirddaddress]{Mine Orlu~\corref{correspondingauthor}}
\ead{m.orlu@ucl.ac.uk}
\author[myfirstaddress]{Marco Ramaioli~\corref{correspondingauthor}}
\ead{m.ramaioli@surrey.ac.uk}
\address[myfirstaddress]{Department of
Chemical and Process Engineering,
Faculty of Engineering and Physical
Sciences, University of Surrey,
Guildford, GU2 7XH, United Kingdom}


\address[mythirddaddress]{School
of Pharmacy, University College
London, London, WC1N 1AX, United Kingdom}

\cortext[correspondingauthor]{Corresponding author. \textit{Tel.:} 0044 1483 68 6514}

\begin{abstract}
Flexible dosing and ease of swallowing are key factors when designing oral drug delivery systems for paediatric and geriatric populations. Multi-particulate oral dosage forms can offer significant benefits over conventional capsules and tablets. 
This study proposes the use of an \textit{in vitro} model to quantitatively investigate the swallowing dynamics in presence of multi-particulates. \textit{In vitro} results were compared against sensory tests that considered the attributes of ease of swallowing and post-swallow residues. Water and hydrocolloids were considered as suspending vehicles, while the suspended phase consisted of cellulose pellets of two different average sizes. Both \textit{in vivo} and \textit{in vitro} tests reported easier swallow for smaller multi-particulates. Besides, water thin liquids appeared not optimal for complete oral clearance of the solids.
The sensory study did not highlight significant differences between the levels of thickness of the hydrocolloids. Conversely, more discriminant results were obtained from \textit{in vitro} tests, suggesting that a minimum critical viscosity is necessary to enable a smooth swallow, but increasing too much the carrier concentration affects swallowing negatively.
These results highlight the important interplay of particle size and suspending vehicle rheology and the meaningful contribution that \textit{in vitro} methods can provide to pre-screening multi-particulate oral drug delivery systems before sensory evaluation.
\end{abstract}

\begin{keyword}
Swallowing \sep Solid oral dosage forms \sep Multi-particulate \sep Flexible dosing \sep Vehicle rheology
\end{keyword}

\end{frontmatter}


\section{Introduction}
\label{sec: Introduction}
Oral solid dosage forms have an enormous cost-saving potential compared to liquid formulations. However, size and shape of classical solid formulations, such as tablets and capsules, can significantly condition patient compliance to a prescribed drug therapy \cite{Lajoinie2014}. Multi-particulates can be designed in a range of particle size offering dose flexibility while retaining the advantages of conventional solid oral dosage forms, such as favourable stability profile \cite{Liu2015}. Moreover, multi-particulate formulations have potential to be better tolerated by patients than conventional tablets \cite{Hayakawa2016,Lopez2015,Mistry2017}. However, further research is needed to shed light on the optimal way of administering multi-particulates to patients: acceptability of multi-particulate administration potentially depends on palatability and rheology of the suspending liquid carrier \cite{Lopez2016}. Thicker fluids are known to ease swallowing of tablets and capsules \cite{Marconati2018,Mistry2017,Kluk2015}.
The rheological properties of a pharmaceutical liquid vehicle require special attention as these will have an impact on several critical quality attributes, such as suspendability, palatability (including appearance, taste and mouth-feel), ease of swallowing and drug bioavailability. In this regard, viscosity modifiers or thickening agents are an essential excipient in the development of liquid vehicles for the administration of medicines. Example of hydrocolloids frequently used as thickening agents include xanthan gum and cellulose derivatives like methylcellulose, hydroxypropyl cellulose and carboxymethyl cellulose \cite{Saha2010}. 
The link between rheology and texture or mouth-feel perception has been extensively studied (especially in food products), although there are still many unknowns given the multifactorial nature of texture perception and palatability \cite{Stokes2013}. Similarly, the correlation between rheology and swallowing is also not fully understood, given the complexity of the physiological process. Thicker liquids have been reported to facilitate swallowing and reduce the risk of bolus penetration-aspiration in patients with swallowing disorders. Nonetheless, there is insufficient evidence to support precise delineation of viscosity boundaries related to such clinical outcomes \cite{Steele2015a}. 
Viscosity measurements taken at shear rates of 50 reciprocal seconds have been found to correlate with initial thickness perception and are commonly used as a reference value to compare between samples \cite{Stokes2013,Chen2012c,Chen2009}. However, higher shear rates, function of bolus rheology, have been indicated for the pharyngeal and esophageal phases of swallowing \cite{Preciado-Mendez2017,Misra2012a}. 

It is important to determine the amount of oral and pharyngeal residues to ease swallowing and ensure thoughtful delivery of the drug. In this context, \textit{in vitro} models of swallowing have recently become available and could be used to understand the end user acceptability of oral dosage forms and to complement the findings obtained from the human panel studies. \textit{In vitro} models make pertinent simplifications of the swallowing physiology and allow understanding the effect of the physical properties of a bolus on its flow \cite{Qazi2017c,Marconati2018b} elucidating their mechanism of action. This study investigated the use of an \textit{in vitro} model of oral cavity -originally developed for studying the oral phase of swallowing of liquids- to evaluate different liquid carriers as a suspending vehicle to administer multi-particulate formulations.

\section{Materials and Methods}
\label{Materials and Methods}


Cellulose pellets (Cellets\textsuperscript{\textregistered} 200 and Cellets\textsuperscript{\textregistered} 700) provided by Pharmatrans Sanaq AG (Basel, Switzerland) were considered as model multi-particulates. Particle size and sphericity were measured in triplicates by dynamic image analysis using a QICPIC/R02 (Sympatec GmbH, Clausthal-Zellerfeld, Germany). Particle density was measured by gas pycnometry using a MicroMeritics 1305 multi-volume pycnometer (Micromeritics Instrument Corp., Norcross, USA). Results from optical and physical measurements are listed in Table~\ref{Cellets}.

Water and two hydrocolloids were considered as suspending vehicles. Xanthan gum (`XG', Xantural 180) was supplied by CP Kelco (Leatherhead, UK) and sodium carboxymethyl-cellulose (`CMC', Blanose 7HF-PH) was provided by Ashland (Covington, USA). The aqueous solutions of XG and CMC were prepared in DI water at room temperature. The study considered three different solutions concentration per each of the two hydrocolloid: 0,25, 0,5 and 1\% for the gum and 0,5, 1 and 1,5\% for the carboxylmethyl cellulose salt. For sensory evaluation, a 0.1\% vanillin was added to mask any potential taste and smell of the polymers. The flavouring ingredient was supplied by Sigma-Aldrich (Irvine, UK).

\subsection{Rheological characterization}
Shear tests of the liquid samples were preformed at 22°C using a HAAKE RheoStress 600 rheometer ( 	Thermo Fisher Scientific, Waltham, USA)  equipped with a cone and plate geometry (d=60 mm $\alpha$=2°). 
Flow curve were run in triplicates in the range of shear rates between 0.01 and 1000 reciprocal seconds. 



Sedimentation of cellulose pellets in the aqueous vehicles was quantified by time taken for an homogeneous suspension to completely clarify the top 1/3 of a 50 ml filled sample holder (d=30 h=115 mm) provided by TPP Techno Plastic Products (Trasadingen, Switzerland). The concentration of Cellets (250 mg of solids in 3 ml of vehicle) was kept consistent throughout the sedimentation tests and equal to that used during the sensory evaluation and the \textit{in vitro} tests. 

\subsection{The \textit{in vitro} swallowing model}
\begin{figure}
\centering
\includegraphics[scale=0.10]{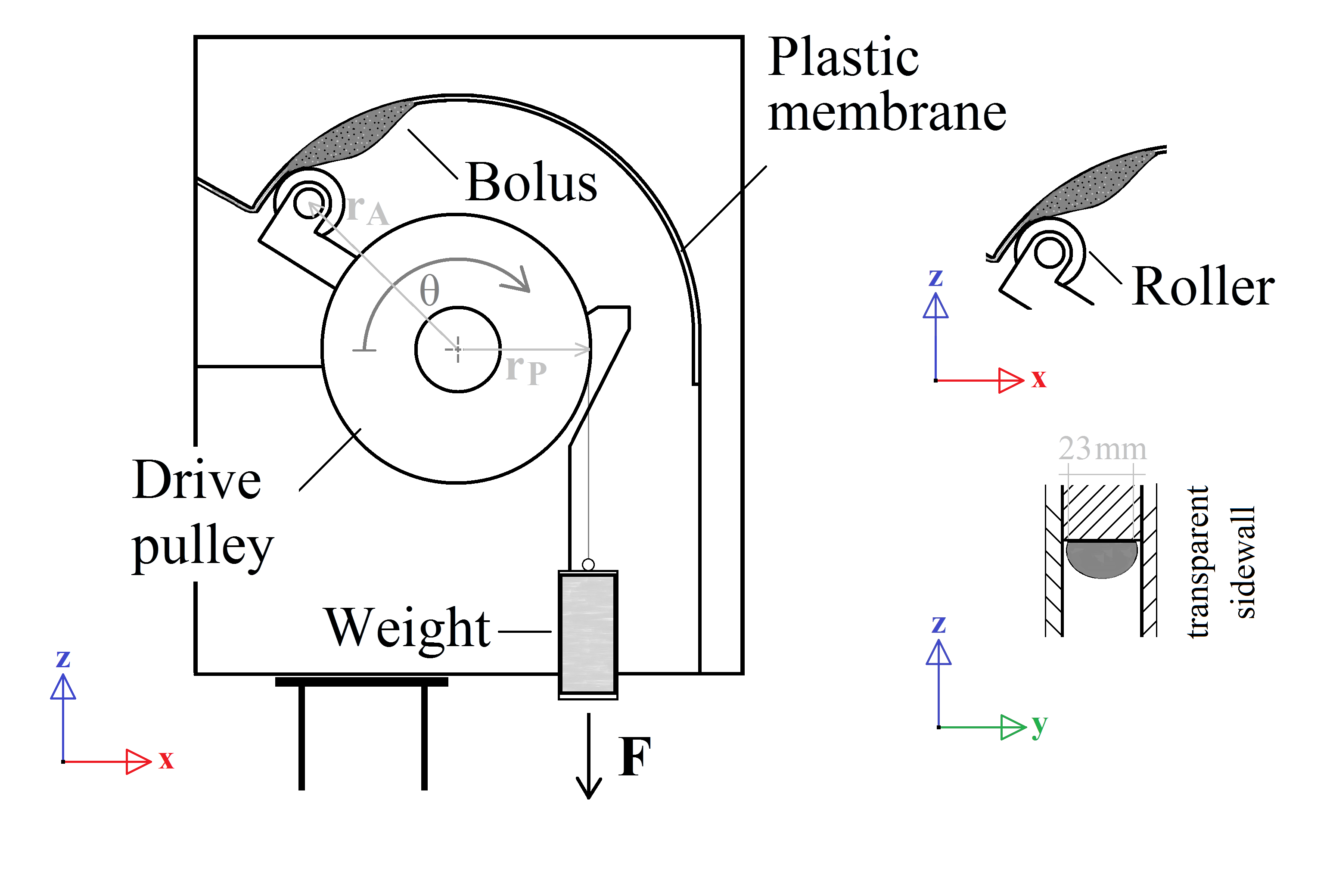}
\caption{\textcolor{black}{Schematics of the \textit{in vitro} setup.}}
\label{Setup}
\end{figure}

The experimental setup used to study the oral phase of swallowing is illustrated in (Fig.~\ref{Setup}). This model simplifies the \textit{in vivo} flow pattern considering its bi-dimensional projection on the sagittal plane. A thin, flat and compliant membrane is glued below a rigid surface mimicking the human palate, and used to constrain and hold the bolus. Cellets were first suspended in the carrier liquid and the suspension was then manually injected into the membrane through its anterior opening (Fig.~\ref{Setup}). The propulsion of the bolus was generated by a roller, sealing anteriorly the membrane filled with the suspension. To mimic the tongue, an elastomeric roller with a Young modulus of E=25 kPa was used. The modulus was determined with uniaxial compression tests at low strain ($\gamma$=0-15\%).
The thin membrane and the roller provided together the two lingual functions of bolus containment and propulsion.
The roller was supported by a pivoting arm, attached to a revolving shaft driven through a set of hanging weights, as schematically depicted in Fig.~\ref{Setup}. Upon triggering of the experiment, the roller moved, following the curved path, squeezing the liquid bolus through the PE membrane. 

In this study a roller driving force of 2 N was used, corresponding to applied torques of 57 mNm and a generated maximum pressures on the bolus tail of approximately 11 kPa, consistently with \textit{in vivo} data from the literature \cite{Hayoun2015a}. Results from the same \textit{in vitro} model were successfully validated against \textit{in vivo} ultrasound measurements with thickened fluids and the swallowing simulator matched well the \textit{in vivo} bolus dynamics when applying a driving force of 2 N\cite{Mowlavi2016}.

During the experiments with multi-particulates, the roller movement was triggered within 60 seconds from the initial bolus loading to mitigate the effect of particle sedimentation. As the roller propelled the bolus posteriorly, lateral images of the experiment were acquired using a fast camera, model ac1920-155 um, (Basler AG, Ahrensburg, Germany). The recording speed was set at 150 frames per second. This high temporal resolution allowed to precisely measure the \textit{in vitro} oral transit time, defined as the time required to clear the bolus front from the plastic membrane that mimics the oral cavity. This indicator was used to infer the ease of swallowing of the different solutions (Table~\ref{Question}).
After each experiment the amount of liquid and solid residues left in the \textit{in vitro} oral cavity was recorded. Four repeats, in randomized order, were taken per each set of experimental variables to assess the variability and robustness of the \textit{in vitro} setup.

\subsection{\textit{In vivo} tests}
The sensory study comprised a total of 30 volunteers, 9 men (age 22.4 $\pm$ 3.9 years) and 21 women (age 25.2 $\pm$ 4.8 years) who had no history of swallowing difficulties \cite{Lopez2018}. Ethical approval was obtained from UCL Research Ethics Committee (Project ID: 4612-011).

The participants were divided into six subgroups and three sessions were organized for sensory evaluation. In each session, the volunteers were offered the six hydrocolloid solutions, water and vanillin-flavoured water in randomized order.

The particle size of the Cellets was not varied within the same daily session. A control test in absence of multi-particulates was run to assess the baseline score for the different liquid solutions.
Each participant was asked to rate the attribute ease of swallowing using a 5-point hedonic scale (1-extremely easy to 5-extremely difficult) and the attributes presence of oral residue based on a 5-point magnitude scale (1-not perceptible to 5-extremely perceptible). furthermore, anecdotal feedbacks of the participants were also registered. Descriptive statistics are presented as mean±SD. A Kruskal-Wallis test was used to identify statistically significant differences among the set of data collected. Significance was set at p\textless 0.01 with 95\% confidence level.

Samples for sensory tests were prepared individually, by mixing 250 mg of cellulose pellets in of 3 ml of liquid vehicle. The suspension was handed to the volunteers with a spoon.
The same volume of liquid and weight fraction of Cellets used throughout the \textit{in vitro} experimental campaign was also considered during the sensory tests.

\section{Results and discussion}

\begin{figure}
\centering
\includegraphics[scale=0.4]{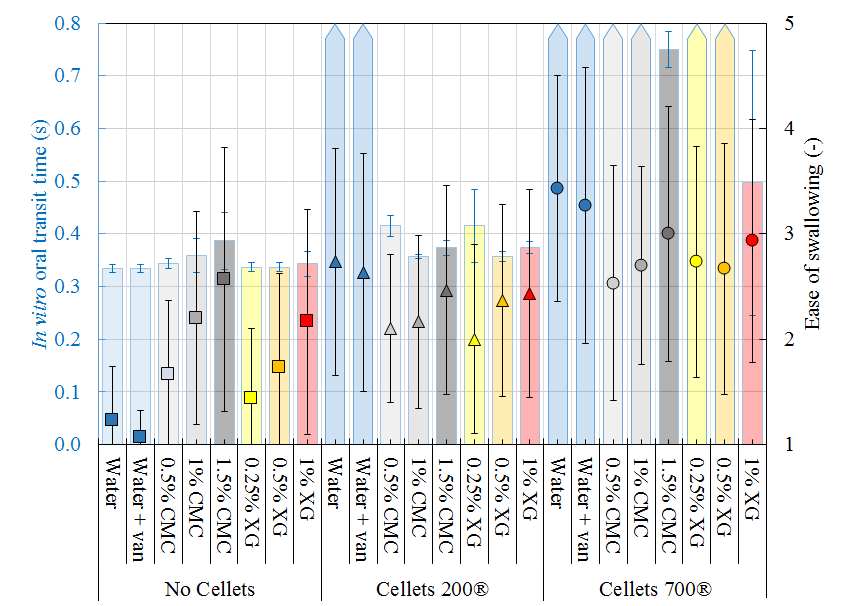}
\caption{\textcolor{black}{\textit{In vitro} oral transit time (bars) and sensory attribute of ease of swallowing (markers).}}
\label{Swallowing}
\end{figure}

The role of suspending vehicle rheology was initially assessed considering control trials in absence of solids. The sensory evaluation showed a noticeable decrease in the attribute of ease of swallowing (p\textless 0.001) moving from water to the hydrocolloids and when the viscosity of the liquid vehicles was increased (Fig.~\ref{Swallowing}). 

\begin{figure}
\centering
\includegraphics[scale=0.4]{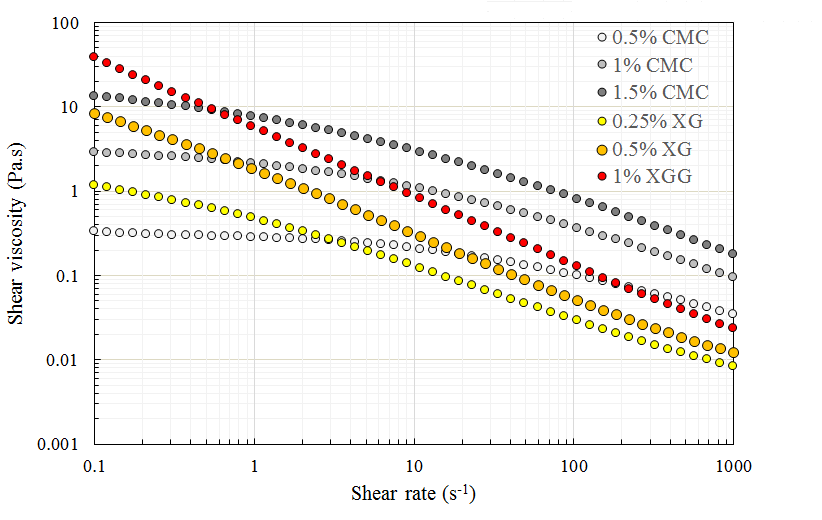}
\caption{Shear viscosity as a function of shear rate for the different hydrocolloids.}
\label{Rheology}
\end{figure}

XG hydrogels were slightly easier to swallow than CMC hydrogels, differences between both sets of hydrogels were statistically significant (p\textless 0.018). An almost linear correlation was found between the average scores given by the volunteers and the concentration of hydrocolloids in water. 
These results are in line with previous publications that report more effortful swallows with increasing hydrocolloid concentration \cite{Hayakawa2014}. Furthermore, a previous study indicates that cellulose-based thickeners were perceived slightly more viscous than XG-based thickeners \cite{Matta2006}. On the other hand, higher slickness scores (i.e. perceived slimy sensation) were also reported for XG hydrogels compared to other cellulose or starch-based thickeners \cite{Matta2006,Hadde2016}. These results are consistent with the pronounced shear thinning behaviour of XG (Fig.~\ref{Rheology}).

In support of the sensory findings, in terms of ease of swallowing, the \textit{in vitro} experiments highlighted an increased duration of the time required for bolus transit moving from water to the hydrogels. A positive correlation between the measured \textit{in vitro} oral transit time and the CMC concentration was observed. Conversely, tests with XG showed no significant variations with respect to concentration (Fig.~\ref{Swallowing}). 

\begin{figure}
\centering
\includegraphics[scale=0.4]{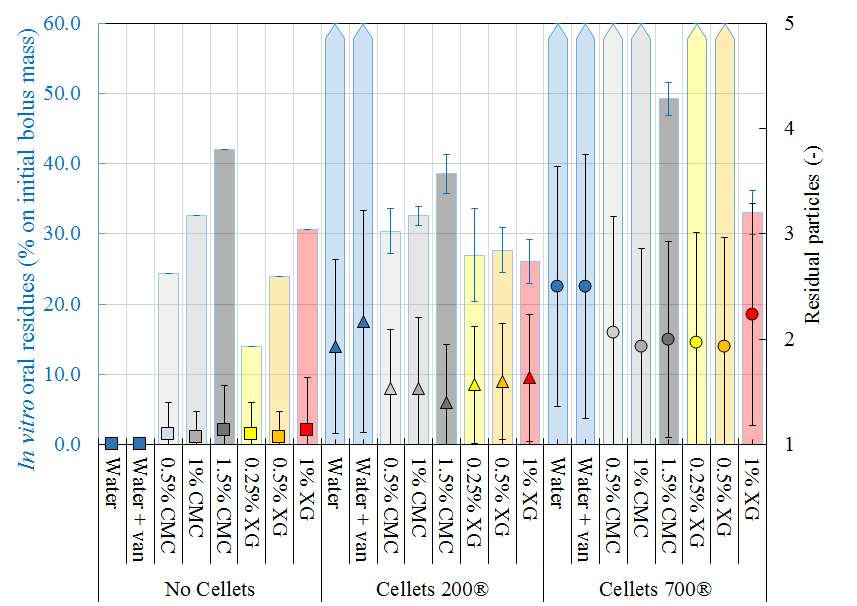}
\caption{\textcolor{black}{Relative amount of \textit{in vitro} post-swallow residues (bars) and sensory scores for residual particles (markers).}}
\label{Residues}
\end{figure}

The healthy volunteers also rated the differences among the liquids in terms of after-swallow feel. Results indicate a minor increase in the amount of residues (p\textless 0.025) when swallowing hydrogels, compared to water. \textit{In vitro}, the amount of oral residues after each experimental run was found to be dependent upon concentration and type of vehicle considered. Water boli left no significant residual mass in the \textit{in vitro} oral cavity whilst a significant portion of the initial bolus mass was not ejected when testing thick vehicles. The amount of residues increased with the concentration of the solutions and was significantly higher for CMC than for XG solutions at the highest concentrations considered (Fig.~\ref{Residues}).
When considering vehicles with similar viscosity in the range of shear rates considered important during swallowing (50-300 reciprocal seconds \cite{Steele2015a}), similar results, in terms of residual mass, were obtained for the thickest gum solution (1\% XG) and the low to intermediate thickness of CMC vehicles.

A good overall correlation between \textit{in vitro} and sensory evaluation results is outlined in terms of increased swallowing difficulty and oral transit time as the viscosity level increased. 
Anecdotal feedbacks collected during the panel tests remark, in a few occasions, the noticeable increase in the swallowing effort required for the thick polymer solutions. In particular, the feedback from 4 volunteers quoted the thickest CMC sample as being very viscous, difficult to swallow and leading to persistent residues in the mouth after swallowing.

Results from the panel tests also show that swallowing of multi-particulates was considered more difficult as the particle size was increased. An average score of 2.36 was attributed to the smaller multi-particulates (Cellets 200), compared to 2.91 for the larger Cellets 700 (p\textless 0.001).
Irrespectively of the size of the particles, multi-particulates dispersed in polymeric hydrogels were easier to swallow by approximately 0.50 points than multi-particulates dispersed in water (p\textless 0.001). 

Despite the contrasting shear thinning behaviour of XG and CMC, the sensory results of the 5-point scales did not indicate a significant difference between the hydrocolloids, that proved effective in facilitating oral delivery of the multi-particulate formulations. Both sets of hydrogels received comparable scores for ease of swallowing (2.52 for XG and 2.49 for the CMC vehicles). However, the average swallowing score for the thickest CMC solutions (1 and 1.5\%) was less significantly modified by the presence of multi-particulates compared to XG.

Aside from worsening the swallowing experience, administration of multi-particulates in water also leads to a gritty sensation followed by a feeling of incomplete clearance of particles from the mouth. 
The feeling of residual particles in the mouth increased with increasing size of the multi-particulates. Panellists rated an average score of 1.67 for Cellets 200 and 2.14 for Cellets 700 (p\textless 0.001). However, the use of polymeric hydrogels reduced the feeling of particles in the mouth after swallowing by approximately 0.5 points on average as compared to water (p\textless 0.001). No significant differences were found in the scores for residual particles between XG and CMC hydrogels. 
Similarly, scores for particle residues were not strongly correlated with the concentration of the polymers. 

\begin{figure}
\centering
\includegraphics[scale=0.4]{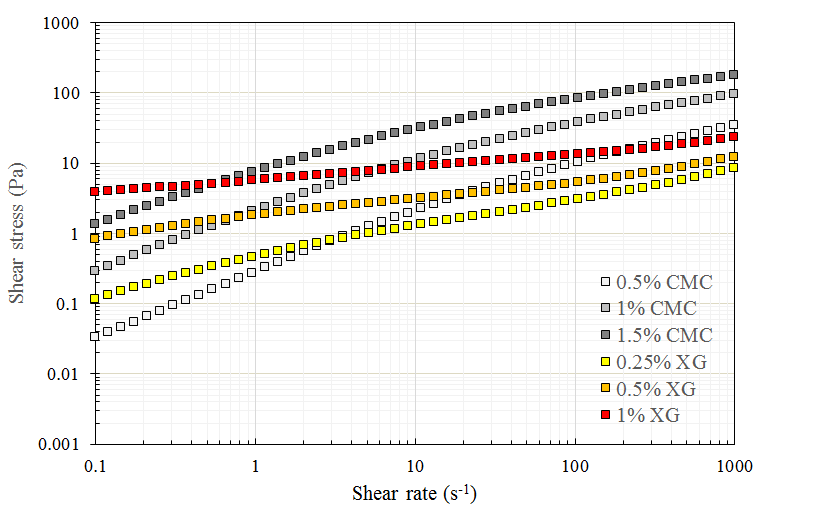}
\caption{Shear stress as a function of shear rate for the different hydrocolloids.}
\label{Stress}
\end{figure}

Anecdotal comments from the volunteers reveal that XG hydrogels were slightly superior multi-particulate carriers than CMC. In presence of Cellets 200, swallowing of the thinnest XG solution was associated to drinking water whilst the 0.5\% XG vehicle was reported viscous, but still easy to swallow and not much persistent in the mouth after the first swallow.
Conversely, some participants reported difficulty when swallowing micro-particulates in CMC. Complaints also came from the higher effort required to clear the mouth from residues. This latter aspect might be linked to the higher stress required to shear CMC at the shear rates relevant to the oral phase of swallowing (Fig.~\ref{Stress}).

The \textit{in vitro} experiments in presence of suspended particles led to a general increase in the measured oral transit time and amount of post swallow residues. Thicker liquid vehicles facilitated \textit{in vitro} swallowing of multi-particulates. \textit{In vitro} experiments also highlighted the importance of particle suspendability. Mechanical jamming (clogging) was observed whenever the suspended particle size was increased and the rheology of the vehicle was unable to prevent a rapid sedimentation. This was always experienced with water for both the particle sizes tested. Hydrocolloids allowed instead a smooth bolus flow with the smallest particles (Cellets 200). However, only the two thickest XG and CMC solutions allowed to run the \textit{in vitro} experiment when the average particle size of the suspended phase was increased to 970 $\mu$m (Cellets 700). Among the two hydrocolloids, faster sedimentation times were registered for CMC (Table \ref{Sedimentation}), consistently with the higher zero shear rate viscosity of XG (Fig.~\ref{Rheology}).

\begin{figure}
\centering
\includegraphics[scale=0.45]{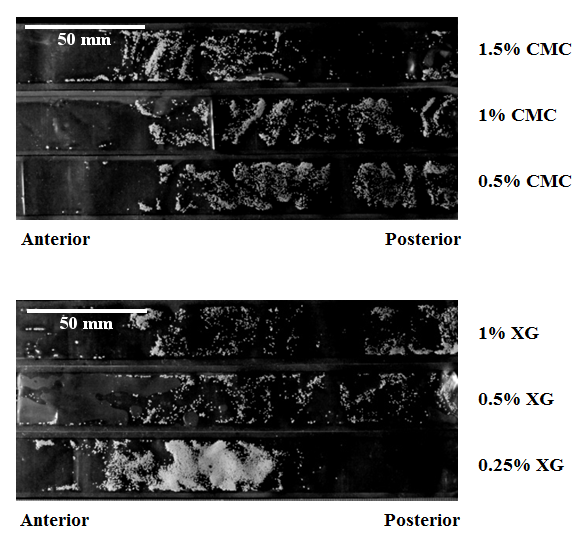}
\caption{\textcolor{black}{Qualitative comparison of the solid residues left after \textit{in vitro} swallowing of Cellets\textsuperscript{\textregistered} 200 in aqueous solutions of XG and CMC.}}
\label{Membranes}
\end{figure}

The \textit{in vitro} results are consistent with the sensory results, since participants stated that larger amounts of Cellets remained in the oral cavity when swallowing samples in thinner vehicles. Some volunteers explicitly mentioned the fact that thin vehicles seemed unable to effectively suspend the solid particles. As a result, solids were left behind the liquid bolus during swallowing. On the other hand, thicker vehicles were perceived as more effective to disperse and transport the intake of multi-particulates.
Consistently with the perceived ease of swallow, \textit{in vitro} transit times increased with increasing particle size. With this respect, XG performs better \textit{in vitro} than CMC, although a smoother bolus flow was observed with CMC. This is reflected in the larger SD observed for XG samples in Fig.~\ref{Swallowing}.
The amount of post-swallow residues (liquid and solid) left in the \textit{in vitro} oral cavity increased with the particle size of the multi-particulates (Fig.~\ref{Residues}). Above the critical viscosity required to avoid particle sedimentation and clogging, the carrier viscosity does not bring a benefit. Visual inspection of the plastic membranes from the \textit{in vitro} experiment seems consistent with the hypothesis that thinner vehicles leave more residual particles behind (Fig.~\ref{Membranes}). Accordingly, this could compensate for the increase in post-swallow residues with the viscosity of the vehicles, observed in absence of suspended particles (Fig.~\ref{Residues}). A positive correlation between the \textit{in vitro} residues (Fig.~\ref{Residues}) and the perceived ease of swallowing (Fig.~\ref{Swallowing}) is also visible when considering the effect of the vehicle concentration, particularly without particles and with the smallest particles. This would be consistent with a higher total residue left in the oral cavity as a result of the perceived increased swallowing difficulty with more concentrated carriers.

\tiny
\begin{longtable}{m{18mm}m{18mm}m{21mm}m{21mm}m{21mm}}
\caption{Particle size and density of the cellulose pellets.}
\label{Cellets}\\
\toprule
\multicolumn{1}{c}{\begin{tabular}[c]{@{}c@{}}Product\\ name \end{tabular}} &
\multicolumn{1}{c}{\begin{tabular}[c]{@{}c@{}}d50\\ (SD n=3)\end{tabular}} &
\multicolumn{1}{c}{\begin{tabular}[c]{@{}c@{}}d90\\ (SD n=3)\end{tabular}} & 
\multicolumn{1}{c}{\begin{tabular}[c]{@{}c@{}}Sphericity\\ (SD n=3)\end{tabular}}
& \multicolumn{1}{c}{\begin{tabular}[c]{@{}c@{}}Density\\ (SD n=3)\end{tabular}} 
\\
\midrule
Cellets\textsuperscript{\textregistered} 200 & 325.3 (0.1) $\mu$m & 358.8 (0.2) $\mu$m & 0.874 (0.003) & 1470 (40) $\text{kg}~\text{m}^{-3}$ \\
\midrule
Cellets\textsuperscript{\textregistered} 700 & 891.2 (0.1) $\mu$m & 970.1 (0.3) $\mu$m & 0.902 (0.002) & 1340 (60) $\text{kg}~\text{m}^{-3}$ \\
\midrule
\end{longtable}
\normalsize

\tiny
\begin{longtable}{m{25mm}m{33mm}m{21mm}m{25mm}}
\caption{\textit{In vivo} and \textit{in vitro} assessment of the swallowing tasks}
\label{Question}\\
\toprule
\multicolumn{1}{c}{\begin{tabular}[c]{@{}c@{}}Sensory \\ attribute \end{tabular}} &
\multicolumn{1}{c}{\begin{tabular}[c]{@{}c@{}}Question \end{tabular}} &
\multicolumn{1}{c}{\begin{tabular}[c]{@{}c@{}}Score \\ range \end{tabular}} &
\multicolumn{1}{c}{\begin{tabular}[c]{@{}c@{}}\textit{in vitro} \\ indicator \end{tabular}}
\\
\midrule
Ease of swallowing & Please rate the ease of swallowinging of the sample & 1 = Extremely easy to 5 = Extremely difficult & Time to bolus front ejection \\
\midrule
Residual particles & After rinsing your mouth with water, can you still feel the ``bits'' in your mouth? & 1 = No bits or imperceptible to 5 = Extremely perceptible & Ratio of post-swallow residues to initial bolus mass \\
\midrule
\end{longtable}
\normalsize

\tiny
\begin{longtable}{m{13mm}m{18mm}m{23mm}}
\caption{Sedimentation time of cellulose pellets in the different liquid vehicles.}
\label{Sedimentation}\\
\toprule
\multicolumn{1}{c}{\begin{tabular}[c]{@{}c@{}}Liquid\\ vehicle \end{tabular}} &
\multicolumn{1}{c}{\begin{tabular}[c]{@{}c@{}}Sedimentation time \\ Cellets\textsuperscript{\textregistered} 200\end{tabular}} &
\multicolumn{1}{c}{\begin{tabular}[c]{@{}c@{}}Sedimentation time \\ Cellets\textsuperscript{\textregistered} 700\end{tabular}}
\\
\midrule
Water & \textless 0.5 min & \textless 0.5 min \\
\midrule
0.5\% CMC &10.5 (0.6) min & 3.34 (0.6) min \\
\midrule
1\% CMC & \textgreater30 min & 22.0 (1.1) min \\
\midrule
1.5\% CMC & \textgreater30.0 min & \textgreater30.0 min \\
\midrule
0.25\% XG & \textgreater30.0 min & 15.5 (0.8) min \\
\midrule
0.5\% XG & \textgreater30.0 min & \textgreater30.0 min \\
\midrule
1.5\% XG & \textgreater30.0 min & \textgreater30.0 min \\
\midrule
\end{longtable}
\normalsize

\section{Conclusions}
Ease of administration and high palatability are key requirements to address the needs of specific populations of patients. 
The study aimed at assessing the ease of swallowing of multi-particulates as a function of their mean particle size and the suspending vehicle rheology.
Both the \textit{in vitro} model and the sensory tests outlined that water-thin vehicles were not optimal for Cellets palatability and oral transport. A critical viscosity threshold for smooth swallowing was observed both \textit{in vivo} and \textit{in vitro}. \textit{In vivo}, above this threshold, differences between vehicles were not significant for healthy volunteers, who rated all samples, on average, on the positive side of the scale. However, analysis of the anecdotal feedback suggested that samples of medium consistency (i.e. 0.50\% XG and 1.00\% CMC) were preferred.
Preference for smaller multi-particulates was also expressed by the volunteers, both in terms of ease of swallowing and lower amount of post-swallow residues. 

The \textit{in vitro} results showed that smaller particles eased bolus transport and reduced residues. Particle suspendability was also a key factor. The residues measured in the \textit{in vitro} test were able to discriminate formulations, even when sensory tests did not indicate a clear difference. In this regard, clear examples were obtained with Cellets 200, both when comparing different CMC concentration and CMC vs XG.
However, a higher critical hydrocolloid concentration seems necessary for a smooth \textit{in vitro} swallow with Cellets 700, when compared to the sensory results.

This confirms the positive contribution that \textit{in vitro} swallowing tests can give to complement sensory studies and gain a solid understanding of the mechanisms involved, while highlighting also some areas for further development.
\textit{In vitro} tests can therefore complement sensory studies in the design of novel formulations.

\section*{Acknowledgments}
MM and MR acknowledge the financial support offered by Nestlé Health Science. FL, CT and MO acknowledge funding from EPSRC for a PhD studentship within the CDT in Targeted Therapeutics and Formulation Sciences (EPSRC grant: EP/I01375X/1).

\bibliographystyle{bibliography}\biboptions{authoryear}

\bibliography{bibliography}

\end{document}